\documentclass[twocolumn,trackchanges]{aastex631}

\shorttitle{A massive quiescent galaxy at $z=4.53$}
\shortauthors{Kakimoto et al.}
\DeclareUnicodeCharacter{2061}{}
\begin{document}

\title{A massive quiescent galaxy in a group environment at $z=4.53$}

\correspondingauthor{Takumi Kakimoto}
\email{takumi.kakimoto@grad.nao.ac.jp}

\author[0000-0003-2918-9890]{Takumi Kakimoto}
\affiliation{Department of Astronomical Science, The Graduate University for Advanced Studies, SOKENDAI, 2-21-1 Osawa, Mitaka, Tokyo 181-8588, Japan}
\affiliation{National Astronomical Observatory of Japan, 2-21-1 Osawa, Mitaka, Tokyo 181-8588, Japan}

\author[0000-0002-5011-5178]{Masayuki Tanaka}
\affiliation{National Astronomical Observatory of Japan, 2-21-1 Osawa, Mitaka, Tokyo 181-8588, Japan}
\affiliation{Department of Astronomical Science, The Graduate University for Advanced Studies, SOKENDAI, 2-21-1 Osawa, Mitaka, Tokyo 181-8588, Japan}


\author[0000-0003-3228-7264]{Masato Onodera}
\affiliation{Subaru Telescope, National Astronomical Observatory of Japan, National Institutes of Natural Sciences (NINS), 650 North A'ohoku Place, Hilo, HI 96720, USA}
\affiliation{Department of Astronomical Science, The Graduate University for Advanced Studies, SOKENDAI, 2-21-1 Osawa, Mitaka, Tokyo 181-8588, Japan}

\author[0000-0003-4442-2750]{Rhythm Shimakawa}
\affiliation{Waseda Institute for Advanced Study (WIAS), Waseda University, 1-21-1, Nishi-Waseda, Shinjuku, Tokyo 169-0051, Japan}
\affiliation{Center for Data Science, Waseda University, 1-6-1, Nishi-Waseda, Shinjuku, Tokyo 169-0051, Japan}

\author[0000-0002-9665-0440]{Po-Feng Wu}
\affiliation{Institute of Astrophysics, National Taiwan University, Taipei 10617, Taiwan}
\affiliation{Department of Physics and Center for Theoretical Physics, National Taiwan University, Taipei 10617, Taiwan}
\affiliation{Physics Division, National Center for Theoretical Sciences, Taipei 10617, Taiwan}

\author[0000-0003-4196-5960]{Katriona M. L. Gould}
\affiliation{Cosmic Dawn Center (DAWN), Denmark}
\affiliation{Niels Bohr Institute, University of Copenhagen, Jagtvej 128, DK-2200 Copenhagen N, Denmark}

\author[0000-0002-9453-0381]{Kei Ito}
\altaffiliation{JSPS Research Fellow (PD)}
\affiliation{Department of Astronomy, School of Science, The University of Tokyo, 7-3-1, Hongo, Bunkyo-ku, Tokyo, 113-0033, Japan}

\author[0000-0002-8412-7951]{Shuowen Jin}
\affiliation{Cosmic Dawn Center (DAWN), Denmark}
\affiliation{DTU Space, Technical University of Denmark, Elektrovej 327, DK-2800 Kgs. Lyngby, Denmark}

\author[0000-0002-7598-5292]{Mariko Kubo}
\affiliation{Astronomical Institute, Tohoku University, Aoba-ku, Sendai 980-8578, Japan}

\author[0000-0002-3560-1346]{Tomoko L. Suzuki}
\affiliation{Kavli Institute for the Physics and Mathematics of the Universe (WPI), The University of Tokyo Institutes for Advanced Study, The University of Tokyo, Kashiwa, Chiba 277-8583, Japan}

\author[0000-0003-3631-7176]{Sune Toft}
\affiliation{Cosmic Dawn Center (DAWN), Denmark}
\affiliation{Niels Bohr Institute, University of Copenhagen, Jagtvej 128, DK-2200 Copenhagen N, Denmark}

\author[0000-0001-6477-4011]{Francesco Valentino}
\affiliation{European Southern Observatory, Karl-Schwarzschild-Str. 2, 85748 Garching, Germany}
\affiliation{Cosmic Dawn Center (DAWN), Denmark}

\author[0000-0001-6229-4858]{Kiyoto Yabe}
\affiliation{Subaru Telescope, National Astronomical Observatory of Japan, National Institutes of Natural Sciences (NINS), 650 North A’ohoku Place, Hilo, HI 96720, USA}



\begin{abstract}
We report on the spectroscopic confirmation of a massive quiescent galaxy at $z_\mathrm{spec}=4.53$ in the COSMOS field. The object was first identified as a galaxy with suppressed star formation at $z_\mathrm{phot}\sim4.65$ from the COSMOS2020 catalog. The follow-up spectroscopy with Keck/MOSFIRE in the $K$-band reveals faint [O\,{\footnotesize II}] emission and the Balmer break, indicative of evolved stellar populations. We perform the spectral energy distribution fitting using photometry and spectrum to infer physical properties. The obtained stellar mass is high ($M_*\sim 10^{10.8}\,M_\odot$) and the current star formation rate is more than 1 dex below that of main-sequence galaxies at $z=4.5$.  
Its star formation history suggests that this galaxy experienced rapid quenching from $z\sim 5$. The galaxy is among the youngest quiescent galaxies confirmed so far at $z_\mathrm{spec}>3$ with $z_\mathrm{form}\sim5.2\ (200\,\mathrm{Myr}$ ago), which is the epoch when 50\% of total stellar mass was formed.
A unique aspect of the galaxy is that it is in an extremely dense region; there are four massive star-forming galaxies at $4.4<z_\mathrm{phot}<4.7$ located within 150 physical kpc from the galaxy. Interestingly, three of them have strongly overlapping virial radii with that of the central quiescent galaxy ($\sim 70\,\mathrm{kpc}$), suggesting that the over-density region is likely the highest redshift candidate of a dense group with a spectroscopically confirmed quiescent galaxy at the center. The group provides us with a unique opportunity to gain insights into the role of the group environment for quenching at $z\sim5$, which corresponds to the formation epoch of massive elliptical galaxies in the local Universe.
\end{abstract}



\section{Introduction} \label{sec:intro}
Massive elliptical galaxies in the local Universe are known to be dominated by old stellar populations with little cold gas and dust, exhibiting no active ongoing star formation (e.g., \citealp{1991ARA&A..29..581Y,Gallazzi_2005,Bellstedt2020}). Since these massive galaxies likely have played a major role in cosmic star formation activities in the early Universe, it is very important to study their formation mechanisms. 
The star formation histories (SFHs) of massive elliptical galaxies inferred from spectroscopic observations show that they experienced an intense burst of star formation in the early Universe and then rapidly quenched (stopped their star formation, \citealp{Nelan_2005,Thomas_2005,Thomas2010,Renzini_2006}). However, the physical drivers of their starburst and subsequent quenching remain unclear, which is one of the most outstanding issues in the field of galaxy formation and evolution \citep{Man2018}.

To elucidate these problems, a lot of efforts have been put into the search for progenitors of local elliptical galaxies at high redshifts and examine them closer to the time of quenching. Over the last two decades, massive quiescent galaxies with no significant ongoing star formation at $z\sim 2$, which are likely the progenitors of local ellipticals, have been extensively studied (e.g., \citealp{Glazebrook2004,Cimatti2004,Cimatti2008,Daddi_2005,Kriek_2009,Kriek2019,Whitaker_2012,Toft_2012,Onodera2012,Onodera2015,Belli_2017,Belli_2019,Morishita2019,Stockmann_2020}). 
These researches infer SFHs of the galaxies and confirm the evidence of starburst at $z>3$ and short quenching timescale.
These galaxies also have very large stellar masses $(M_*>10^{11}\,M_\odot)$ and compact sizes $(R_e<2\,\mathrm{kpc})$ compared to their counterparts in the local Universe (e.g., \citealp{vanDokkum2008,Cassata2011,Newman2012,vanderWel2014,Lustig2021}). 

Thanks to the recent development of sensitive near-infrared (NIR) spectrographs, massive quiescent galaxies at up to $z\sim 4$ are spectroscopically confirmed (e.g., \citealp{Glazebrook2017,Schreiber_2018,Forrest_2020,Forrest_2020b,Valentino_2020,D_Eugenio_2020,Eugenio_2021,nanayakkara_2022,Antwi-Danso_2023,Tanaka2023}). This epoch is consistent with the formation epoch inferred from the local elliptical galaxies \citep{Thomas2010}, and the identification of massive quiescent galaxies at $z>4$ is critical to understand their quenching mechanism. The number of sources confirmed at $z > 4$ is still relatively limited \citep{Tanaka_2019,carnall_2023}, but deep ground-based observations and new JWST spectroscopy are about to change the landscape. Also, recent cosmological simulations have difficulties in reproducing these galaxies (e.g., \citealp{Schreiber_2018,Guarnieri_2019,Cecchi_2019,Merlin_2019,Valentino_2020,Gould_2023}). Observational confirmations of massive quiescent galaxies at $z>4$ are an important step to constrain the galaxy formation models.

In terms of the environment in which galaxies reside, it is known that elliptical galaxies dominate clusters of galaxies in the nearby Universe (e.g., \citealp{1980ApJ...236..351D}). 
This morphology-density relation shows that the surrounding environment may affect physical properties of galaxies. Actually, massive quiescent galaxies dominate high-density environments at $z<1$ (e.g., \citealp{Gomez_2003,Peng_2010,Kawinwanichakij2017}). These environments are basic components of large-scale structures in the nearby Universe, and examining the formation process by observing the high-redshift Universe is crucial. At $z>2$, previous studies confirm ``proto-clusters'' which are over-density regions of galaxies and are considered progenitors of local clusters (e.g., \citealp{Capak2011,Toshikawa_2014,Lemaux_2018,Oteo2018,Harikane2019,Hu2021,Brinch2024}). However, these proto-clusters are often identified using star-forming galaxies as a tracer, and there are only a few systems with (confirmed) quiescent galaxies (e.g., \citealp{Kubo_2021,Kubo_2022,Kalita_2021,McConachie_2022,Ito_2023}). To understand the role of the environment in galaxy quenching at high redshift, it is crucial to examine more (proto-)clusters and look for a sign of environmental effects in the act.

In this paper, we report the confirmation of a massive quiescent galaxy at $z=4.53$ in the COSMOS field using the Keck/MOSFIRE spectrograph. The galaxy is located in a group environment, and we discuss a possible role of environment for quenching at this high redshift for the first time. This paper is structured as follows. First, we introduce the target selection and spectroscopic follow-up observation in Section \ref{sec:obs}. In Section \ref{sec:SEDfit}, physical properties of the quiescent galaxy inferred from spectral energy distribution (SED) fitting are summarized. Then, we discuss the galaxy's surrounding environment as well as its possible formation scenarios in Section \ref{sec:group}. Finally, we conclude the paper in Section \ref{sec:Concl}. We assume a \cite{Chabrier_2003} initial mass function (IMF) and a flat $\mathrm{\Lambda CDM}$ cosmology with $H_0 = 70\,\mathrm{km\,s^{-1}\,Mpc^{-1}}, \Omega_m = 0.3$, and $\Omega_\Lambda = 0.7$. All magnitudes are in the AB system \citep{1983ApJ...266..713O}. 

\section{Observation} \label{sec:obs}
\subsection{Target Selection} \label{subsec:target}
Following the successful confirmation of the $z=4.01$ quiescent galaxy in the Subaru/XMM-Newton Deep Field \citep{Tanaka_2019}, we searched for massive quiescent galaxy candidates at $z>4$ in the COSMOS field \citep{Scoville2007}. The COSMOS field is $\sim 2\,\mathrm{deg^2}$ and has been extensively observed from the X-ray to the radio wavelengths, which allows us to obtain highly accurate photometric redshifts (photo-$z$'s).  As summarized in \cite{Ito_2022}, we applied our photo-$z$ code \citep{Tanaka_2015} to the COSMOS2020 \textsc{Classic} catalog \citep{Weaver_2022}. The code adopts the \cite{BC03} models, assuming exponentially declining star formation histories (SFHs, $SFR(t) \propto e^{-t/\tau}$), solar metallicities, the \cite{Calzetti_2000} dust attenuation curve, and the \cite{Chabrier_2003} IMF. From the comparisons between photo-$z$'s and spectroscopic redshifts from the literature, our photo-$z$'s are confirmed to be accurate for massive quiescent galaxies at $2<z<4$ in the COSMOS field ($\sigma[\Delta z/(1+z_\mathrm{spec})]=0.03, \mathrm{Median}(\Delta z)=0.011$: \citealp{Ito_2022}, and references therein).

Based on this photo-$z$ catalog, we applied the following cuts to select quiescent galaxy candidates; (1) $z_\mathrm{phot}>4$ with little probability of being at lower redshifts to ensure that the Balmer break is in the $K$-band, (2) $M_\mathrm{*,c} \gtrsim 5\times10^{10}\,M_\odot$ so that the target is sufficiently bright for us to trace its continuum ($M_\mathrm{*,c}$ refers to a stellar mass in the photo-$z$ catalog), and (3) star formation rate (SFR) inferred from the SED fitting is more than 1 dex below the star-forming main sequence \citep{Schreiber_2015}. The last criterion is effectively similar to the rest-frame $UVJ$ selection \citep{Williams_2009}.

After careful visual screening, we identify a massive quiescent galaxy candidate at $z_\mathrm{phot}\sim 4.65$ with these cuts. 
Its SED exhibits a strong Balmer break, which suggests that the galaxy experienced an intense burst of star formation in the recent past and that its SFR has been declining since then (post-starburst phase). This object (COSMOS-1047519) is the subject of the paper. We note that there are other quiescent galaxy candidates, but most of them are very faint in the $K_s$-band (too high redshifts where the Balmer break is outside of the $K_s$-band or suffer from the dust extinction). The selected candidate is the brightest candidate after the selection.

\subsection{Spectroscopic follow-up} \label{subsec:follow}
We performed spectroscopic follow-up in the $K$-band using the MOSFIRE spectrograph \citep{McLean2010,McLean2012} on the Keck I telescope. We expected to fully cover the entire Balmer break from the photo-$z$. The target is rather faint with a $K_s$-band magnitude of $23.2\,\mathrm{mag}$, and a long integration was required to identify absorption features. However, the weather did not fully cooperate and we could observe roughly half of the allocated time ($\sim9$ hour on-source exposure). 
The observations were performed on 3 half nights in 2022 March and April. We took $\sim 100$ pairs of the ABBA nodding with a 3$\,$min exposure at each position. We nod $\pm 2''$ along the slit for good sky subtraction. We included two bright stars in the mask to monitor the sky conditions (see below). The mean seeing was around $0''.7$.

We reduced the data in the standard manner using the MOSFIRE Data Reduction Pipeline (DRP)\footnote{\url{https://keck-datareductionpipelines.github.io/MosfireDRP/}} as well as custom-designed code. We reduced each A/B pair using DRP (flat-fielding, wavelength calibration, and background subtraction). 
The flux and the FWHM of the bright stars in each exposure were used to evaluate the observing conditions. We used the weight of flux/FWHM in the coaddition of the spectra to give more weight to exposures taken under better conditions. Four exposures were taken through thick clouds and were thus excluded from the coadd.

We observed the A0V star (HD50046) as a spectrophotometric standard star on the first night of the observation. We applied the flux calibration and correction for the telluric absorption by comparing the alpha-Lyr spectrum from HST/CALSPEC \citep{Bohlin_2014} and the observed standard star spectrum. The CALSPEC spectrum was scaled to the standard star using the $K_s$-band photometry from 2MASS \citep{2MASS}.

We performed the optimal extraction \citep{Horne1986} to extract the 1D spectra from the 2D spectra. The spatial profile of the bright star spectrum was fit by Gaussian, and it was used to extract the flux of the target galaxy. The error spectrum was estimated using the standard deviation of the background pixels in the same slit. Finally, the flux of the 1D spectrum is scaled to match the observed photometry to correct for the slit loss.

\begin{figure*}
    \plotone{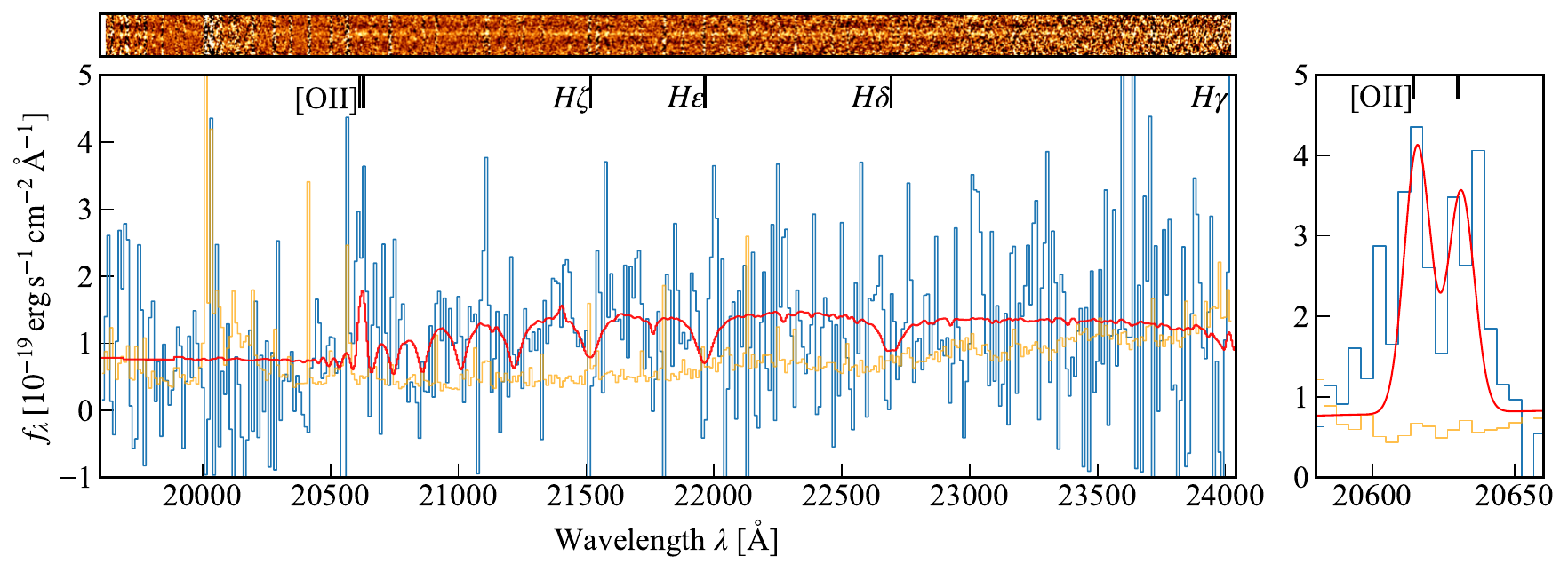}
    \caption{Left: Keck/MOSFIRE $K$-band spectrum of COSMOS-1047519 smoothed over 5 pixels (10.8\,\AA). The top panel shows the two-dimensional spectrum, and the bottom panel shows the one-dimensional spectrum (the blue line) and the associated noise spectrum (the yellow line). The red line shows the best-fit model spectrum from \texttt{prospector} using spectrum and photometry (see Section \ref{sec:SEDfit}). Right: Blow-up of [O\,{\footnotesize II}] doublet lines smoothed over 2 pixels (4.3\,\AA). The red line represents the best-fit Gaussians.}
    \label{fig:spec}
\end{figure*}

\subsection{Redshift confirmation} \label{sec:redshift}
Figure \ref{fig:spec} shows the observed 2D and 1D spectra re-binned over 5 pixels (10.8\,\AA). Due to the low S/N of the spectrum, we could not securely identify absorption lines. For example, the median S/N at $21000\,\mathrm{\AA} < \lambda < 22000\,\mathrm{\AA}$ is only 1.4 per $2.2\,\mathrm{\AA}$ (original resolution). However, the spectrum exhibits a faint [O\,{\footnotesize II}]$\,\lambda\lambda3727,3729$ emission with $\mathrm{S/N_{peak}}\sim 5$. A blow-up of [O\,{\footnotesize II}] is shown in the right panel of Figure \ref{fig:spec}. 
This [O\,{\footnotesize II}] emission line provides a secure spectroscopic redshift for this object. We perform a Monte-Carlo run to fit the [O\,{\footnotesize II}] doublet with two Gaussians and adopt the median as a point estimate and the 68\% interval as an uncertainty, $z_\mathrm{spec}=4.5313 \pm 0.0005$.  This redshift is consistent with the photo-$z$ ($z_\mathrm{phot}= 4.654^{+0.134}_{-0.104}$).  Based on this redshift estimate, we discuss detailed properties of the galaxy in the next Section.

\section{Inferred physical properties} \label{sec:SEDfit}
\subsection{SED Fitting} \label{subsec:SED}
\begin{figure}
    \plotone{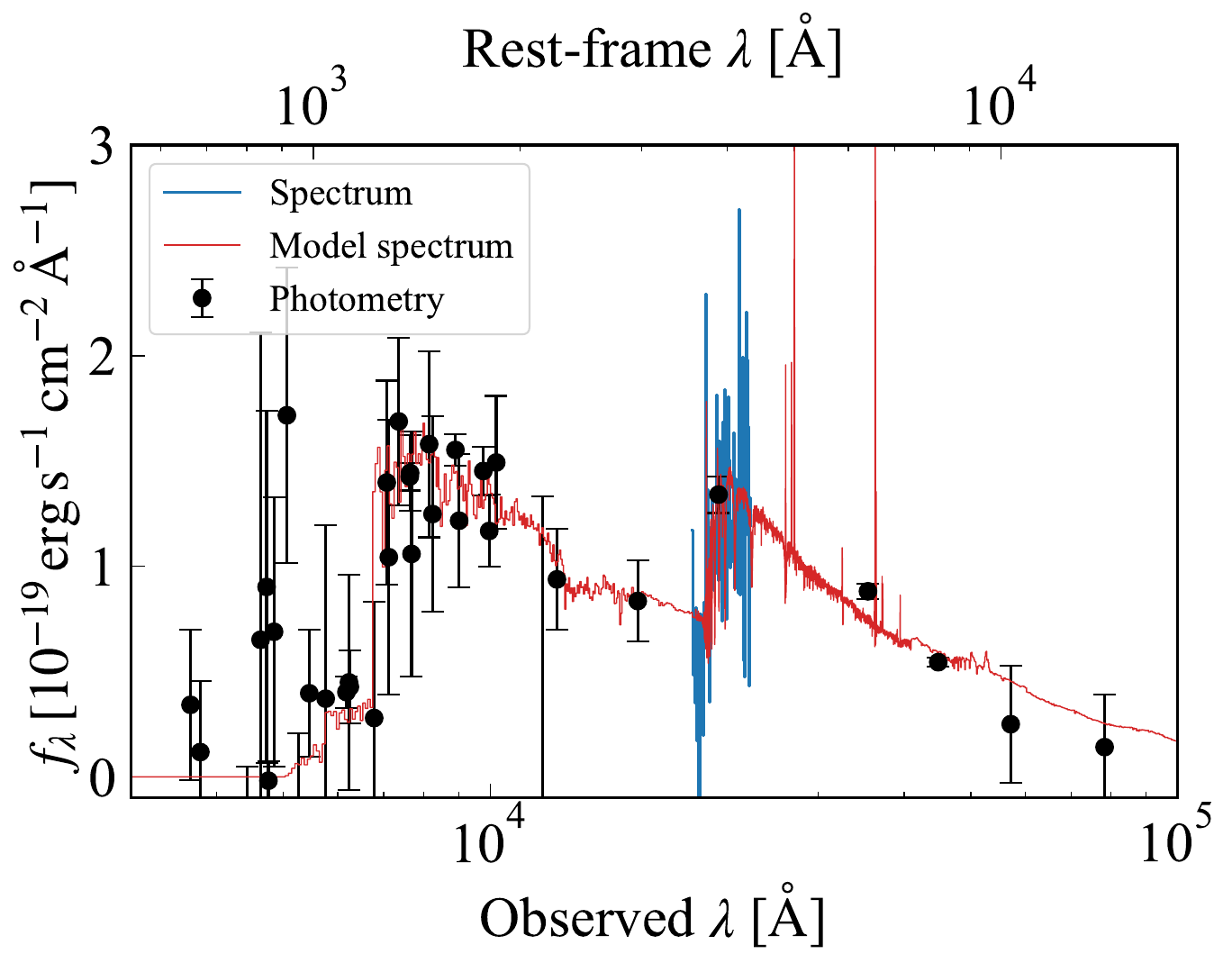} 
    \caption{Spectral energy distribution of the target. The black points and error bars show the photometry from the COSMOS2020 catalog \citep{Weaver_2022}, and the blue spectrum at $\sim 2\,\mathrm{\mu m}$ is the MOSFIRE spectrum. The red spectrum is the best-fit model spectrum from \texttt{prospector}.}
    \label{fig:SED}
\end{figure}

Figure \ref{fig:SED} shows the SED of the galaxy. The black points are the photometry from the COSMOS2020 \textsc{Classic} catalog with 40 bands from $u$-band to IRAC ch4 \citep{Weaver_2022}. The blue spectrum is the MOSIFRE spectrum, which shows the prominent Balmer break, suggesting the post-starburst nature of this galaxy. The spectrum fully covers the break feature as expected, and we can accurately infer its star formation history as shown below. 

We fit the observed SED with \texttt{prospector} \citep{Johnson_2021}. The MOSIFRE spectrum is re-binned over 3 pixels to gain S/N (e.g., the median $\mathrm{S/N} \sim 2.5$ per $6.5\,\mathrm{\AA}$ at $21000\,\mathrm{\AA} < \lambda < 22000\,$\AA). The \texttt{prospector} code uses the Flexible Stellar Population Synthesis model \citep{Conroy_2009,Conroy_2010} to infer physical properties of the galaxy. We assume the \cite{Chabrier_2003} IMF, solar metallicities, the \cite{Madau_1995} IGM absorption model, and the \cite{Calzetti_2000} attenuation curve. The redshift is fixed to the spectroscopic redshift in the fitting.  As we detected the weak [O\,{\footnotesize II}] line, we include nebular emission in the model (fixed the ionization parameter: $U=0.01$).

We assume two models of star formation history (SFH). One is a parametric SFH model with a linear rise of SFR followed by an exponential decrease (``delayed tau-model'', $SFR(t) \propto te^{-t/\tau}$). The parameter $\tau$ represents decay timescale of star formation. We choose this model because it is widely used and can compare with other quiescent galaxies from the literature. 
The other model is a non-parametric SFH model, which allows the SFR in each age bin to vary, thus gives more flexibility in the SFH modeling. There are several priors to use this non-parametric model, and we use ``log-mass prior'' and ``Dirichlet prior'' \citep{Leja_2017,Leja_2019}. The log-mass prior is the most flexible model, but it could lead to a non-physical SFH such as causing intense burst star formation just after the Big Bang. 
Thus, we also use the Dirichlet prior, which forces the SFH to be smoother. As discussed below, the estimated SFHs are similar regardless of the SFH model used.

For all models, we apply the Markov Chain Monte Carlo (MCMC) using \texttt{emcee} \citep{2013PASP..125..306F} to infer the best-fitting parameters and their associated errors. We use the log-uniform prior to the input parameters such as $\tau$, age, and stellar mass. To be specific, in the parametric SFH model, the parameters are allowed to vary between $0.001\,\mathrm{Gyr}<\tau<10\,\mathrm{Gyr}$, $0.13\,\mathrm{Gyr}<age<1.30\,\mathrm{Gyr}$, which is the age of the Universe at $z=4.53$, and $10^{9}\,M_\odot<M_*<10^{12}\,M_\odot$. For dust optical depth, we use the top-hat prior between $0<A_V<6$. In the non-parametric SFH model, we use the uniform prior to constrain the stellar masses of each age-bin. 

The best-fit model spectrum is shown in red in Figs. \ref{fig:spec} and \ref{fig:SED}. The model clearly shows the strong Balmer break in the $K$-band, which indicates that the galaxy recently experienced a starburst and is now in the post-starburst phase. There is relatively strong rest-frame UV continuum, implying that there are residual star formation activities in the galaxy. This is thus likely a young post-starburst and is likely being quenched.  We will further discuss this point using physical properties inferred from the SED fit in the following subsection.

\begin{deluxetable*}{cc|lcccccc}
\tablecaption{Inferred physical properties of the quiescent galaxy (COSMOS-1047519) \label{tab:param}}
\tablehead{
\colhead{$z_\mathrm{phot}$} & \colhead{$z_\mathrm{spec}$} & \colhead{Model} & \colhead{$\log{M_*/M_\odot}$} & \colhead{$SFR_\mathrm{SED}$} & \colhead{$A_V$} & \colhead{$z_\mathrm{form}$} & \colhead{$\tau$} & \colhead{$SFR_\mathrm{[O\,{\footnotesize II}]}$} \\
\colhead{} & \colhead{} & \colhead{} & \colhead{} & \colhead{[$M_\odot\,\mathrm{yr^{-1}}$]}& \colhead{[$\mathrm{mag}$]} & \colhead{} & \colhead{[$\mathrm{Myr}$]} & \colhead{[$M_\odot\,\mathrm{yr^{-1}}$]} 
}
\colnumbers
\startdata 
{       } & {       } & {Delayed tau} & $10.71\pm 0.04$ & $15.1^{+6.5}_{-5.2}$ & $0.74^{+0.08}_{-0.09}$ & $5.01^{+0.11}_{-0.08}$ & $28.4^{+8.4}_{-6.0}$ & $19.3^{+4.2}_{-3.2}$\\ 
$4.654^{+0.134}_{-0.104}$ & $4.5313\pm 0.0005$ & {Log mass} & $10.77\pm 0.05$& $11.2^{+8.5}_{-7.8}$ & $0.69\pm 0.12$ & $5.21^{+0.54}_{-0.28}$ & -- & $18.0^{+4.5}_{-3.5}$\\
{       } & {       } & {Dirichlet} & $10.79\pm 0.01$ & $8.76^{+6.85}_{-4.81}$ & $0.61^{+0.13}_{-0.15}$ & $5.36^{+0.38}_{-0.18}$ & -- & $16.0^{+4.5}_{-3.5}$\\
\enddata

\tablecomments{
\begin{enumerate}
    \setlength{\itemsep}{-2pt}
    \item Photometric redshift inferred from \texttt{MIZUKI} \citep{Tanaka_2015}.
    \item Spectroscopic redshift from the Keck/MOSFIRE spectrum.
    \item Name of the star formation history model used in SED fitting \citep{Johnson_2021}.
    \item Inferred stellar mass.
    \item Star formation rate from the SED fitting averaged over $10^{7.5}\,\mathrm{yr}$.
    \item $V$-band magnitude of dust attenuation from SED fitting.
    \item Formation redshift at which the galaxy formed 50\% of total stellar mass. Here, the lookback times from the SFH are converted into redshifts assuming the adopted cosmology (see Section \ref{sec:intro}).
    \item Quenching timescale only in the delayed tau-model.
    \item Star formation rate from the flux of the [O\,{\footnotesize II}] emission corrected for dust attenuation.
\end{enumerate}
}
\end{deluxetable*}

\subsection{Inferred physical properties and SFH} \label{subsec:property}
From the SED fit, we infer physical properties of the galaxy such as stellar mass, star formation rate (SFR), dust optical depth, and star formation history. These parameters are summarized in Table \ref{tab:param}.  We estimate the SFR from the SED fits and also from [O\,{\footnotesize II}] fit using two Gaussians.  We use the following equation to convert the [O\,{\footnotesize II}] emission flux to the SFR \citep{Kewley_2004,Schreiber_2018}.
\begin{equation} \label{eq:OII}
    SFR_\mathrm{[O\,{\footnotesize II}]}\,[M_\odot\,\mathrm{yr^{-1}}] = 1.59 \times 10^{-8} L_\mathrm{[O\,{\footnotesize II}]}\,[L_\odot].
\end{equation}
We correct for dust attenuation of $L_\mathrm{[O\,{\footnotesize II}]}$ using the dust optical depth from the best-fit model.

\begin{figure}
    \plotone{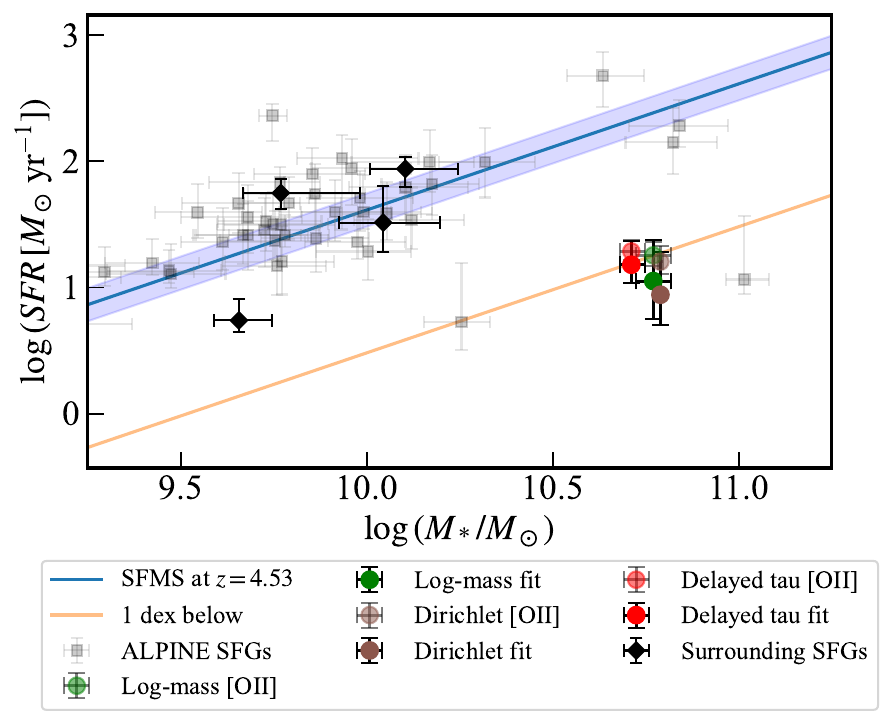} 
    \caption{Stellar mass and SFR relation inferred from the three SFH models (color-coded symbols) compared to the star-forming main sequence (the blue line: \citealp{Schreiber_2015}). The orange line shows the star-forming main sequence shifted by -1 dex. Inferred SFRs from both SED and [O\,{\footnotesize II}] emission corrected for dust are located around or below the orange line. The gray points are the normal star-forming galaxies at $z_\mathrm{spec} \sim 4.5$ from the ALPINE survey \citep{LeFevre2020,Bethermin2020,Faisst_2020}. There are two ALPINE galaxies below the orange line, but the one at $\log{(M_*/M_\odot)} \sim 10.3$ has a contaminating galaxy nearby, and the other one is possibly undergoing dust-obscured star formation \citep{Faisst_2020}. We also include massive star forming galaxies around the quiescent galaxy (the black diamonds, see discussion in Section \ref{sec:group}).}
    \label{fig:mainseq}
\end{figure}

Figure \ref{fig:mainseq} shows the location of the object on a SFR vs. stellar mass diagram. We also plot the normal star-forming galaxies from ALPINE survey  \citep{LeFevre2020,Bethermin2020,Faisst_2020} for reference. The galaxy is massive ($M_* \approx 6.0 \times 10^{10}\,M_\odot$), and has a very low SFR ($SFR \approx 10\,M_\odot\,\mathrm{yr^{-1}}$) averaged over $10^{7.5}\,\mathrm{yr}$. The estimated SFR is more than 1 dex below the star-forming main sequence \citep{Schreiber_2015}. This supports our argument in the previous subsection that the galaxy is a massive quiescent galaxy based on \cite{Schreiber_2018} criteria. If we perform the SED fitting without using the spectrum, uncertainties of physical properties are about 2 times larger. In addition, the height of the Balmer break cannot be fully determined by $K_s$-band photometry only, since the $K_s$-bend is on the break. The spectrum, on the other hand, can firmly determine the break strength, resulting in a more precise age determination. Figure \ref{fig:mainseq} also shows that the [O\,{\footnotesize II}]-based SFRs are higher than those obtained by SED fitting, which we interpret as a contribution from active galactic nuclei (AGN) as we discuss in Section \ref{subsec:AGN} (the model spectrum cannot reproduce [O\,{\footnotesize II}] either in Figure \ref{fig:spec}).

\begin{figure}
    \plotone{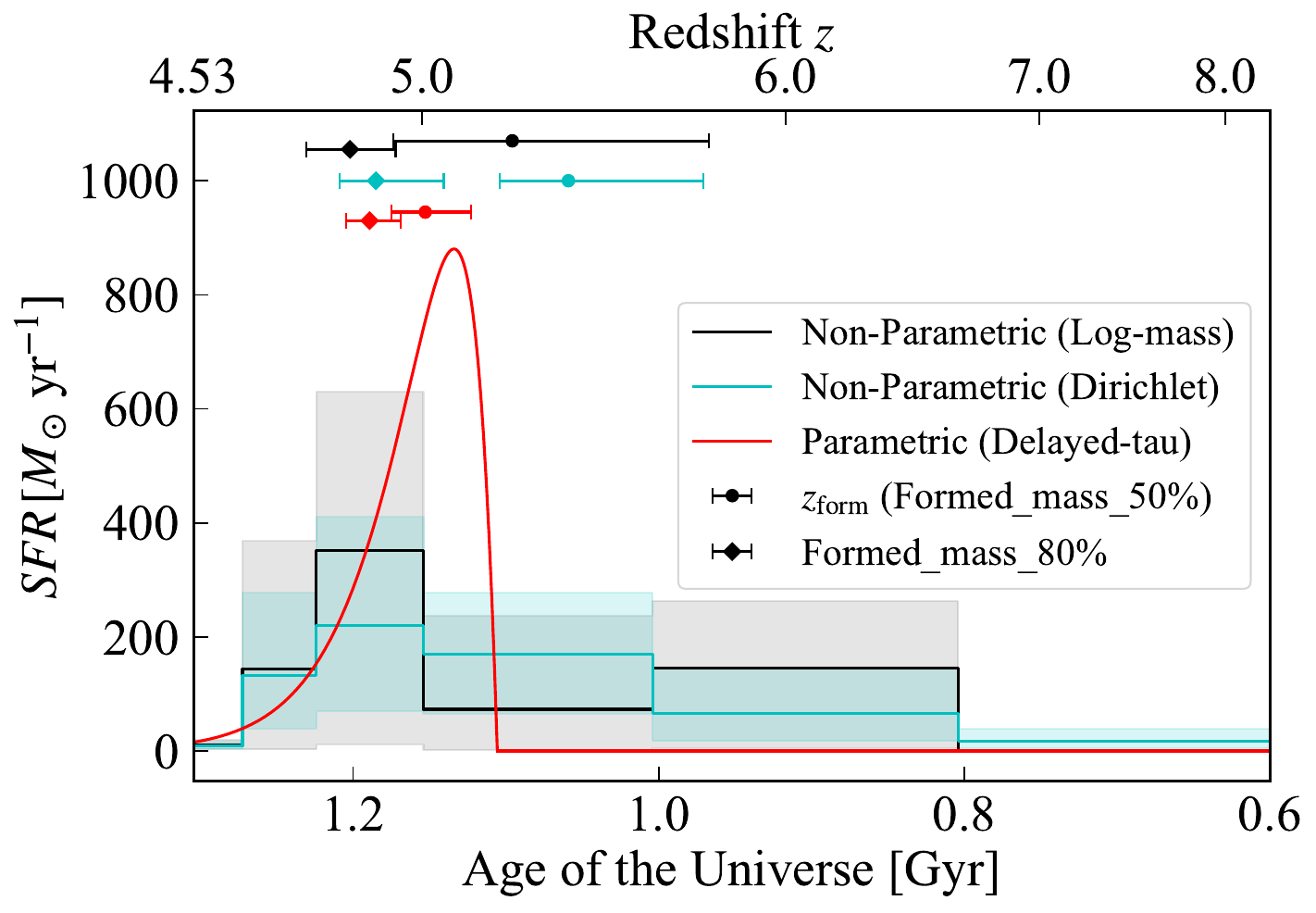}
    \caption{Inferred SFHs of the target; the black, blue, and red lines are the non-parametric SFH model with log-mass prior, non-parametric model with Dirichlet prior, and the delayed-tau model, respectively. The filled circles and diamonds with the error bars indicate the epochs when the galaxy formed 50\% and 80\% of the stars, respectively (the colors correspond to the SFH models).}
    \label{fig:SFH}
\end{figure}

Figure \ref{fig:SFH} shows the inferred star formation history of the galaxy. All of the SFH models are broadly consistent with each other particularly during the quenching epoch; the galaxy experienced a rapid quenching process beginning around $z\sim 5$, which closely aligns with the SFH of massive elliptical galaxies \citep{Thomas2010}. The galaxy is young with a stellar age of $198^{+44}_{-32}\,\mathrm{Myr}$ from delayed tau-model, and the observed low SFR is consistent with the SFHs. Inferred physical parameters and SFH are not significantly changed by the age bins of the non-parametric SFH model. We conclude that this is a massive quiescent (or being quenched) galaxy and is likely a progenitor of massive elliptical galaxies in the local Universe.

The quiescence of the galaxy is also supported by the multi-wavelength data. We look at the infrared to radio wavelengths to see a sign of dust-obscured star formation activity. However, the object is not detected in FIR ($S/N_\mathrm{IR}\sim1.25$, $SFR_\mathrm{IR}<300\,M_\odot\,\mathrm{yr^{-1}}$ in ``Super-deblended'' catalog: \citealp{Jin_2018}) and radio ($f_\mathrm{10\,\mathrm{cm}}=6.3\times 10^{-4}\pm 2.8\,\mathrm{\mu Jy}$ in VLA 3GHz: \citealp{smolcic_2017}). The object has not been observed by ALMA, and we do not have a strong constraint on dust emission at this point. 

\subsection{Comparison to massive quiescent galaxies at $z>3$} \label{subsec:previous}
An interesting aspect of our quiescent galaxy is that it is currently experiencing ongoing suppression of star formation. The prominent Balmer break suggests an evolved stellar population, while the remaining UV continuum as well as the weak [O\,{\footnotesize II}] emission suggests that the galaxy is not completely quenched yet (Figs. \ref{fig:spec} and \ref{fig:SED}). The galaxy has a young luminosity-weighted age, and we must be observing the galaxy close to its primary formation epoch. 

\begin{figure}
    \plotone{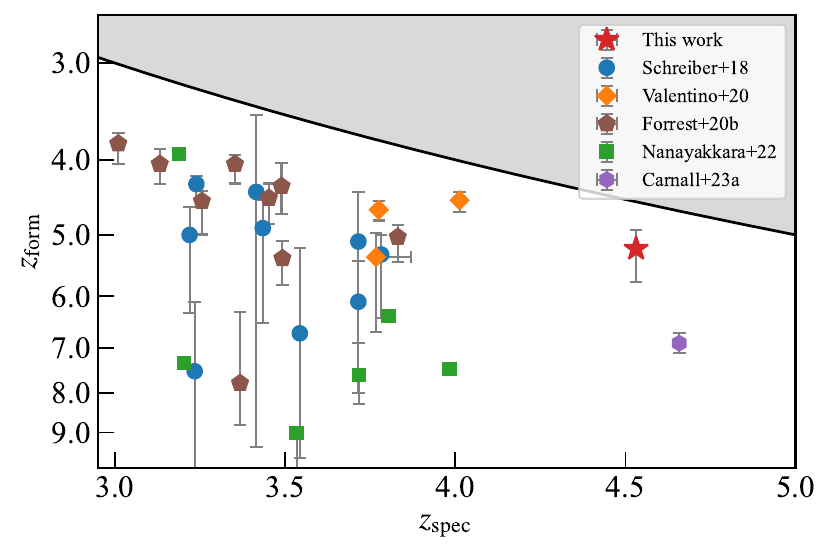} 
    \caption{Relation between formation redshift ($z_\mathrm{form}$) at which the galaxy formed 50\% of total stellar mass and observed redshift ($z_\mathrm{spec}$). We plot COSMOS-1047519 (the red star) and spectroscopic samples of massive quiescent galaxies at $z>3$ from previous research \citep{Schreiber_2018,Valentino_2020,Forrest_2020b,nanayakkara_2022,carnall_2023}. \cite{carnall_2023} use a mass-weighted mean formation time to define $z_\mathrm{form}$.}
    \label{fig:formz}
\end{figure}

Figure \ref{fig:formz} shows the relationship between the formation redshift $(z_\mathrm{form})$ and observed redshift $(z_\mathrm{spec})$ of massive quiescent galaxies at $z_\mathrm{spec}>3$ collected from the literature. The formation redshift is defined by the epoch at which the galaxy formed 50\% of total stellar mass. This figure shows that massive quiescent galaxies at $z>3$ have a wide range of formation epochs (also confirmed at $z<3$: summarized in \citealp{Kalita_2021,Tacchella_2022}). Our galaxy is young and its formation redshift is close to the observed redshift. In addition, its formation redshift ($z_\mathrm{form}\sim5.2$) coincides with the inferred formation epoch of massive elliptical galaxies in the local Universe \citep{Thomas2010}. We thus speculate that the galaxy is a forming (or just formed) progenitor of nearby massive ellipticals and a detailed study of the galaxy may provide us with a clue to the quenching physics.
The other massive quiescent galaxies in the figure are generally older and more time has elapsed between the formation and observed epochs.
For instance, \cite{carnall_2023} confirm a massive quiescent galaxy at $z_\mathrm{spec}= 4.658$. Compared to this galaxy, COSMOS-1047519 is younger and has a higher SFR and larger dust attenuation. However, the stellar mass and the peak SFR in its SFH are similar.  
Therefore, our galaxy is similar to the \cite{carnall_2023} galaxy but we are closer to the quenching epoch.

The young age of our galaxy is likely due to a selection effect; we intentionally targeted the brightest candidate in the $K_s$-band (which covers the Balmer break), and thus we tend to select galaxies with a prominent Balmer break, leading us to young post-starbursts. It is interesting that old quiescent galaxies show little dust attenuation ($A_V\sim 0$: e.g., \citealp{carnall_2023,nanayakkara_2022}), while young ones suffer from a larger amount of attenuation ($A_V\sim 0.7$: This work, \citealp{Valentino_2020}). This indicates that post-starburst galaxies have residual dust.

Based on the non-parametric star formation history (Figure \ref{fig:SFH}), the peak SFR of COSMOS-1047519 is $SFR\sim 200\mathrm{-}400\,M_\odot\,\mathrm{yr^{-1}}$. This is about 2-3 times higher than the SFR of star-forming main sequence galaxies at $z=5$ \citep{Schreiber_2015}. If the galaxy experienced an intense starburst only once, starbursting galaxies are possible progenitors as suggested by previous research (e.g., \citealp{Forrest_2020,Forrest_2020b,Valentino_2020,carnall_2023}). Massive quiescent galaxies at $z\sim 2$ show similar properties to submillimeter galaxies (SMGs) at higher redshifts in terms of masses, effective radii, and number densities \citep{Toft_2014,Gomez-Guijarro2018}. SMGs tend to have a larger star formation rate than the star-forming main sequence. By comparing the physical properties, the peak SFR in SFH of our galaxy is consistent with that of SMGs which have similar stellar mass at $z\sim 5$ (e.g., \citealp{Cunha2015,Miettinen2017}). Moreover, the quenching timescale for our quiescent galaxies is roughly 100\,Myr, which is comparable to the gas depletion timescales of SMGs with similar specific star formation rates at $1<z<5$ \citep{Gomez-Guijarro2022}. This quenching timescale is defined as the period from when the galaxy reaches its peak SFR in the nonparametric SFH until the SFR decreases to 10\% of that peak. Also, \cite{Zavala_2022} summarize the redshift evolution of gas depletion timescale of both starburst galaxies and normal star-forming galaxies and show that the timescale is shorter at the high-redshift Universe. Our galaxy's quenching timescale is consistent with the starburst galaxies' gas depletion timescale at $z\sim 5$.
Thus, our work supports the scenario in which SMGs evolve into quiescent galaxies.

\subsection{Comparisons with simulations and the role of AGNs}\label{subsec:AGN}
The Illustris TNG300 simulation \citep{pellepich2018,2019ComAC...6....2N} is a large cosmological hydrodynamical simulation and reproduces the cosmic evolution from the early Universe to the local Universe. Thus, the simulation is a very useful tool to help us understand how massive galaxies form in the early Universe. We search for similar objects to COSMOS-1047519 in a snapshot at $z=4.43$. However, there are no massive ($M_*>10^{10.5}\,M_\odot$) quiescent galaxies with $sSFR<10^{-9.5}\,\mathrm{yr^{-1}}$. \cite{Hartley2023} confirm the first quiescent galaxy in the TNG300 is at $z\sim 4.2$ although the definition of the quiescent galaxy is slightly different from ours ($sSFR<10^{-10}\,\mathrm{yr^{-1}}$). They conclude that kinetic feedback from AGN has a significant effect on the quenching process of quiescent galaxies reconstructed at $z\sim 4.2$. We note that COSMOS-1047519 is not detected in the deep X-ray observation ($L_\mathrm{2-10\,keV}<5.1\times 10^{43}\,\mathrm{erg\,s^{-1}}$ in Chandra COSMOS Legacy: \citealp{Civano_2016}), but this upper limit of X-ray luminosity does not rule out the possibility of low-luminosity AGN in this galaxy.

Figure \ref{fig:spec} shows that the model spectrum cannot fully reproduce the [O\,{\footnotesize II}] flux. Table \ref{tab:param} and Figure \ref{fig:mainseq} also show that [O\,{\footnotesize II}]-based SFR is higher than that from SED fitting. 
The observed excess suggests that AGN may reside in the galaxy because [O\,{\footnotesize II}] is sensitive to ionization by (low-luminosity) AGN as well as star formation (e.g., \citealp{Yan_2006}). 
\cite{Maseda2021} find that more than half of massive quiescent galaxies at $z<1$ exhibit [O\,{\footnotesize II}] lines and conclude that this feature is due to AGN activities.  At $z>2$, there are massive quiescent galaxies with AGN (e.g., \citealp{Kriek_2009,Marsan_2015,Schreiber_2018,Kubo_2022}). Our galaxy may be a similar case, although it is located at a higher redshift. Future deep X-ray or JWST NIR observations of AGN activity (confirming [O\,{\footnotesize III}], [N\,{\footnotesize II}], and Balmer lines) are necessary to identify or rule out low-luminosity AGN.

\section{Group environment around the quiescent galaxy} \label{sec:group}

\subsection{Over-density in the surrounding structure} 
\begin{figure*}
    \plotone{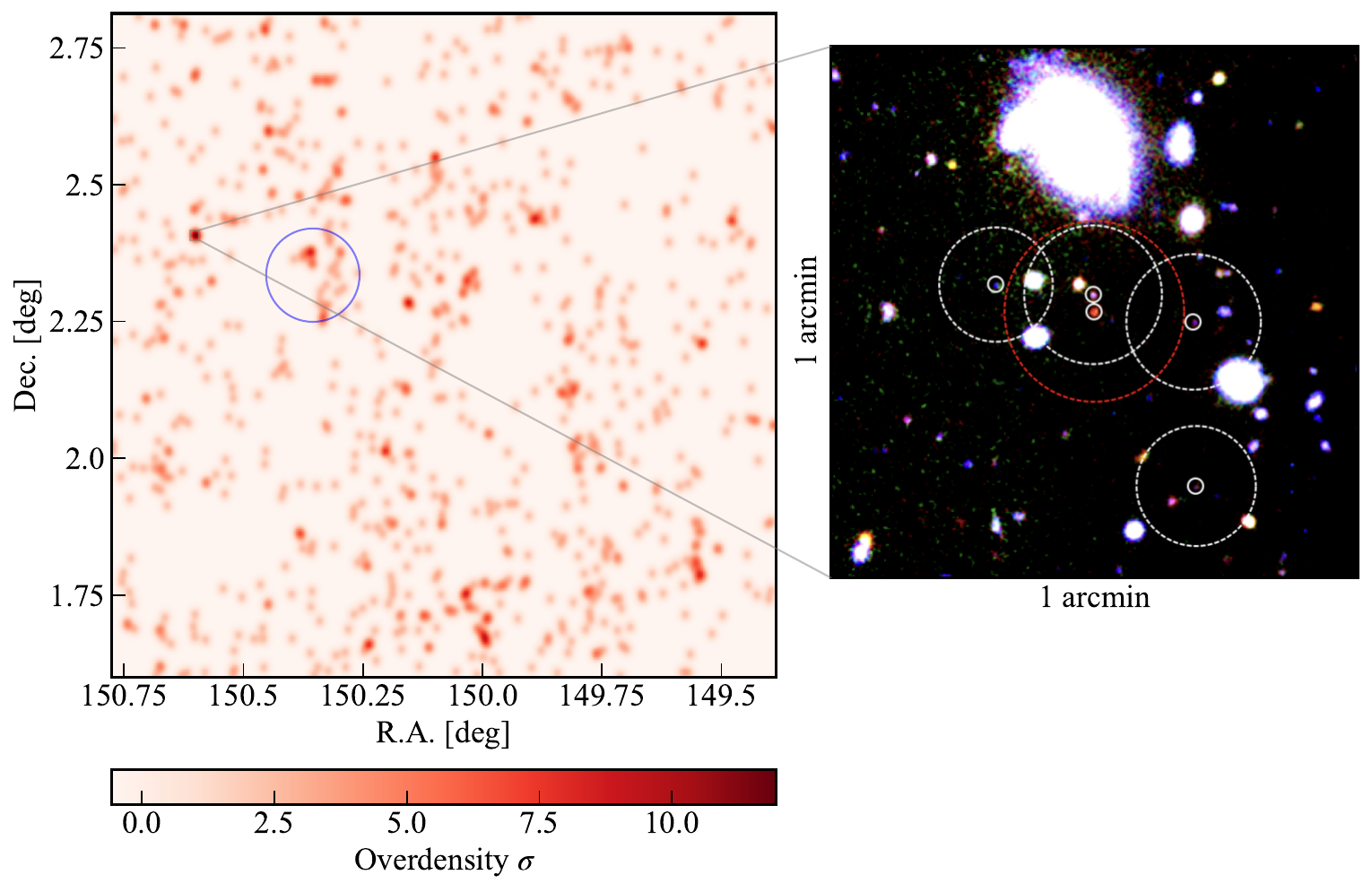} 
    \caption{Distribution of massive ($M_\mathrm{*,c}>10^{10}\,M_\odot$ inferred from the photo-$z$ catalog) galaxies at $4.4<z_\mathrm{phot}<4.7$ in the entire COSMOS field. The red color shows the density excess of the massive galaxy candidates smoothed over $30''$ which corresponds to the scale of the galaxy group candidate using kernel density estimation. The blue circle shows the massive proto-cluster at $z_\mathrm{spec}\sim 4.57$ ($R_\mathrm{proj}=2\,\mathrm{Mpc}$: \citealp{Lemaux_2018}). The right panel shows a pseudo-color image of the area around the quiescent galaxy (Red: VISTA/$K_s$-band, Green: VISTA/$H$-band, Blue: HSC/$i$-band, \citealp{McCracken_2012,Tanaka_2022}). The five galaxies indicated with the white solid circles are the member candidates of this group centered on the quiescent galaxy. Our massive quiescent galaxy at $z=4.53$ (the red point in the RGB image) exists in the densest region ($12\sigma$ in the COSMOS field). The red and white dotted circles show the estimated virial radii of the quiescent galaxy and star-forming member candidates, respectively. The virial radii are estimated based on the stellar masses inferred from \texttt{prospector}.}
    \label{fig:2}
\end{figure*}

\begin{figure*}[tb]
    \centering
    \begin{minipage}{0.49\textwidth}
        \centering
        \includegraphics[width=0.95\textwidth]{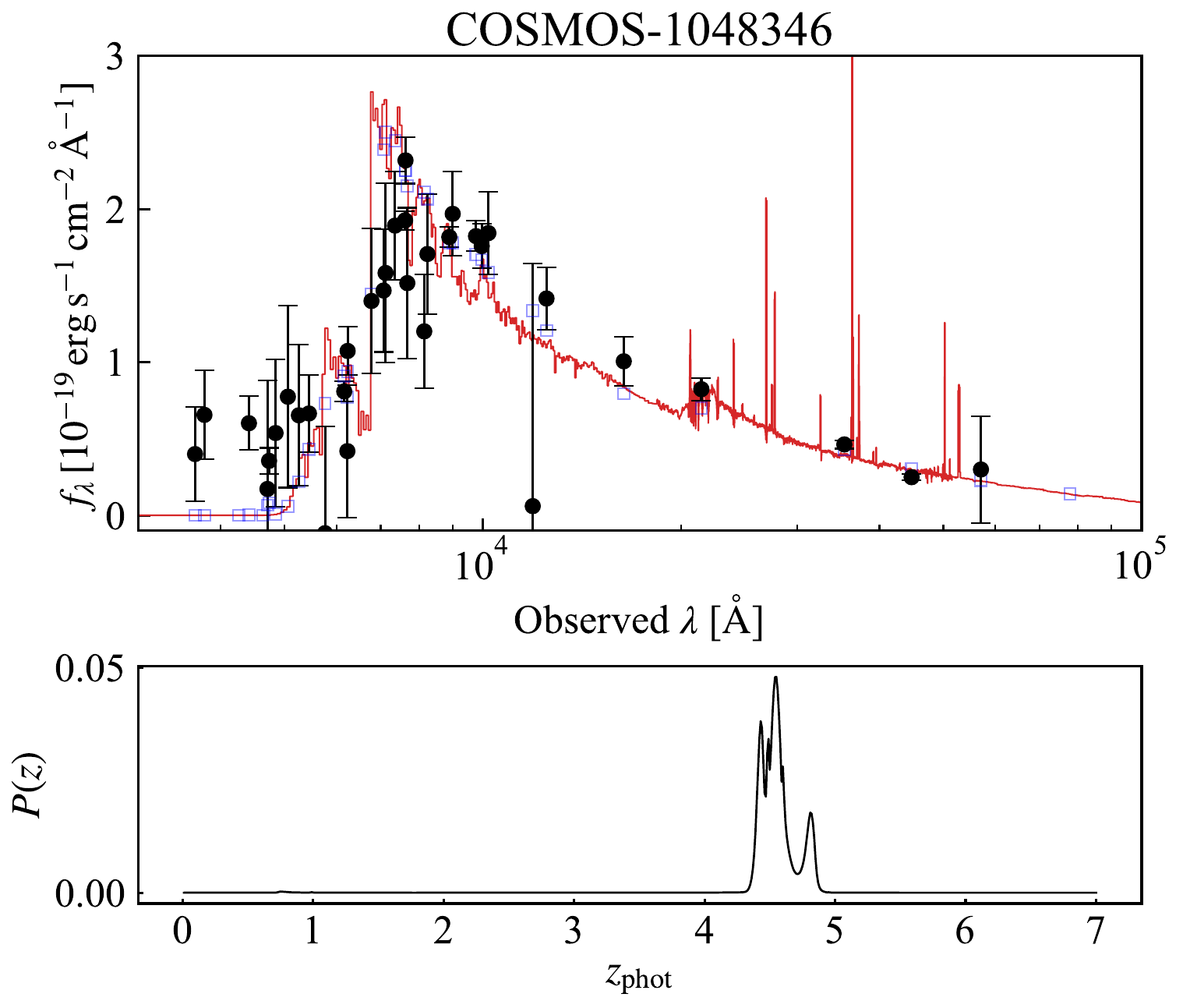}
    \end{minipage}
    \begin{minipage}{0.49\textwidth}
        \centering
        \includegraphics[width=0.95\textwidth]{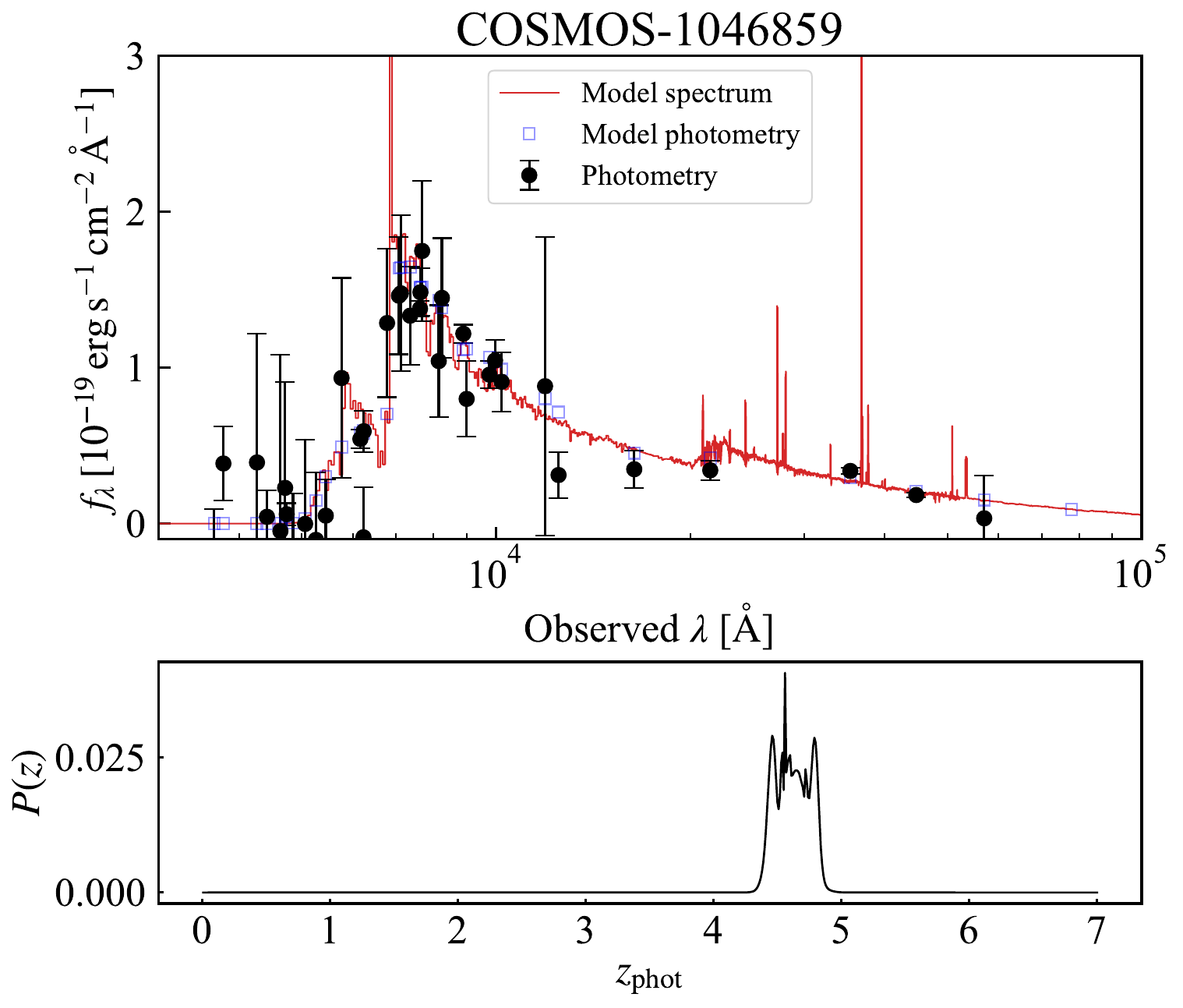}
    \end{minipage}
    \begin{minipage}{0.49\textwidth}
        \centering
        \includegraphics[width=0.95\textwidth]{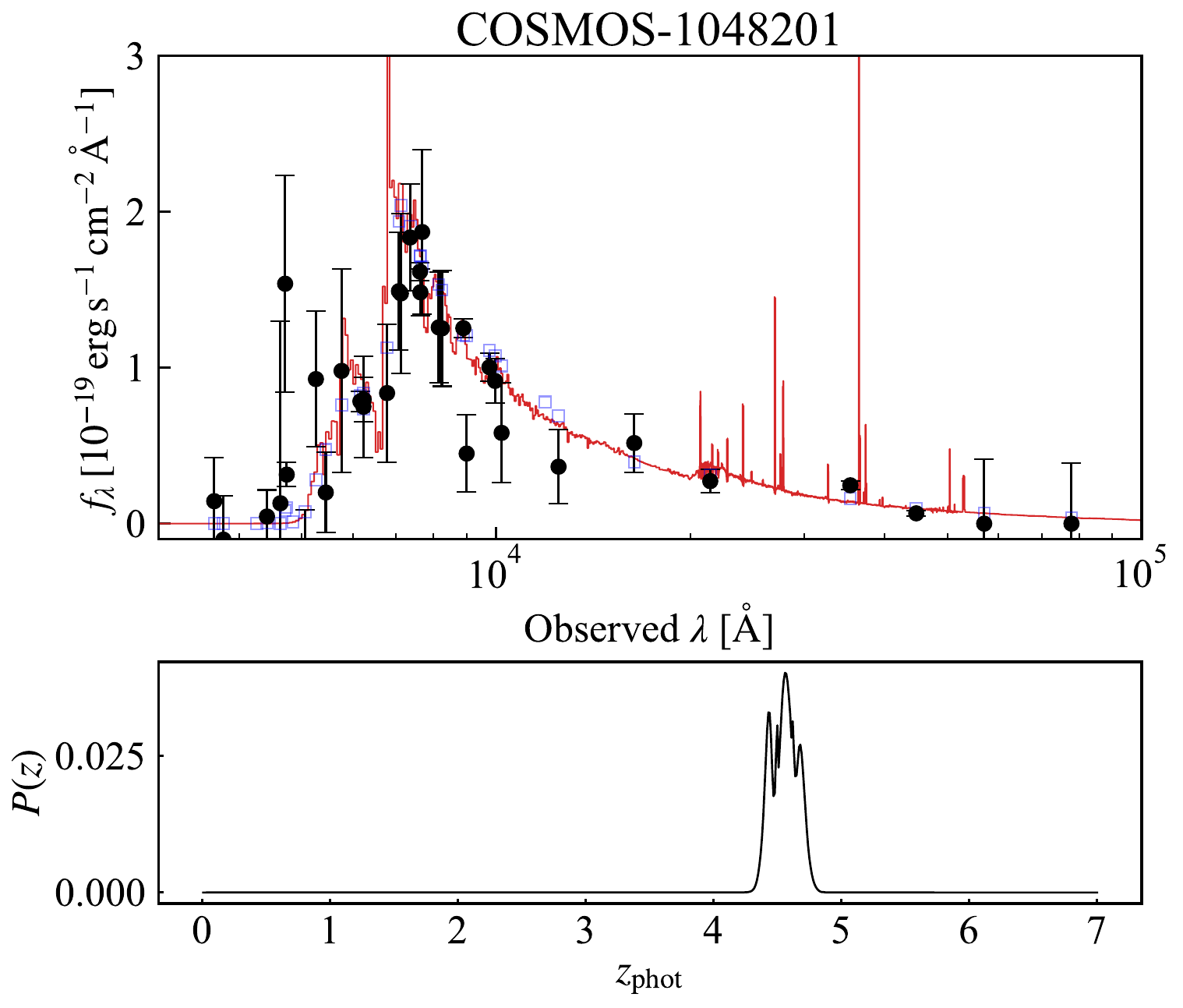}
    \end{minipage}
    \begin{minipage}{0.49\textwidth}
        \centering
        \includegraphics[width=0.95\textwidth]{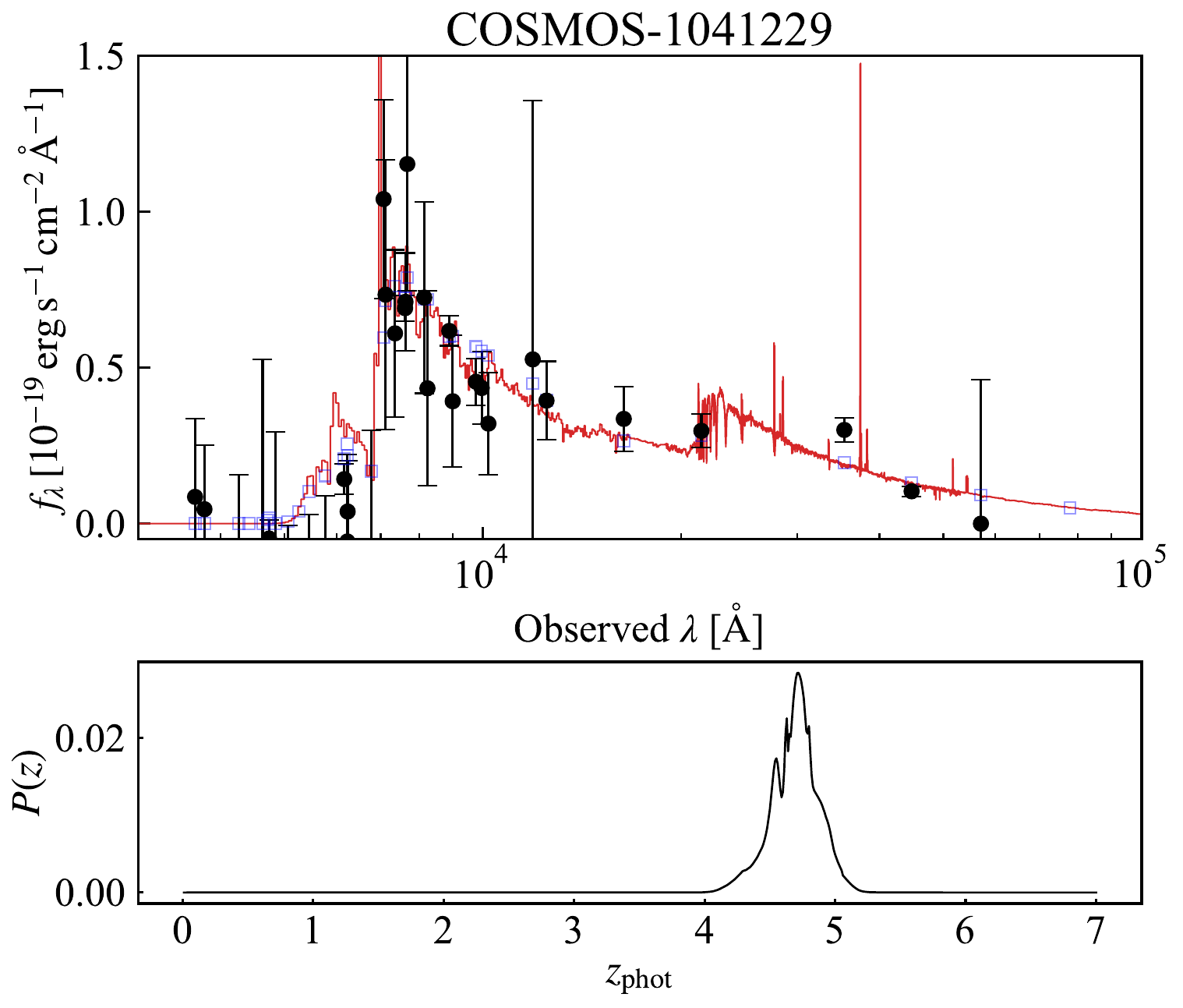}
    \end{minipage}
    \caption{SEDs of the group member candidates. The black points show the photometry from the COSMOS2020 catalog \citep{Weaver_2022}, and the red lines and blue squares indicate the model SEDs from \texttt{MIZUKI} \citep{Tanaka_2015}. The bottom panels show the probability distribution of photometric redshift. All galaxies have a large probability at $z_\mathrm{phot}\sim 4.5$.}
    \label{fig:SEDgroup}
\end{figure*}

To gain deeper insights into quenching physics, we investigate the environment of our galaxy.
We select massive ($M_\mathrm{*,c}>10^{10}\,M_\odot$) galaxies regardless of SFRs at $4.4<z_\mathrm{phot}<4.7$ from the photo-$z$ catalog in the COSMOS field (Section \ref{subsec:target}: \citealp{Ito_2022}). 
The inset of Figure \ref{fig:2} shows a blow-up around the quiescent galaxy (QG).  There is a companion galaxy $(z_\mathrm{phot}=4.535^{+0.168}_{-0.105})$ right next to it with a separation of $2''$. Furthermore, there are 3 massive star-forming galaxies (SFGs) $(z_\mathrm{phot}=4.57\mathrm{-}4.69)$ within $23''$ or 151\,kpc in physical scale. We recall that the quiescent galaxy has a photo-$z$ of 4.65. We summarize the SEDs and probability distributions of photo-$z$ inferred from \texttt{MIZUKI} \citep{Tanaka_2015} in Figure \ref{fig:SEDgroup}, and the coordinates in J2000 are summarized in Table \ref{tab:RADEC}. Each SED shows that all member candidates have bluer SEDs compared to the quiescent galaxy. The probability distributions show that all galaxies have the highest probability at $z_\mathrm{phot}\sim 4.5$, and most of them have consistent photometric redshifts within $1\sigma$ of the quiescent galaxy. These photo-$z$'s are also consistent with the COSMOS2020 catalog.

To evaluate the physical properties of these SFGs and the QG homogeneously, we perform the SED fitting analysis of 4 SFGs by \texttt{prospector} using a parametric SFH model (delayed-tau model with the same priors as those used for the QG). The inferred physical properties are summarized in Table \ref{tab:group}. The SFGs are normal star-forming galaxies (their locations on a SFR vs. stellar mass diagram are shown in Figure \ref{fig:mainseq} as black diamonds) but their stellar masses are smaller than our original estimates from our photo-$z$ code. Yet, these values are still satisfied the 70\% completeness limit of SFGs at $z=4.5$ ($M_*>10^{9.37}\,M_\odot$, \citealp{Weaver_2022complete}). The quiescent galaxy has the largest stellar mass, and this environment may be the highest redshift galaxy group with a spectroscopically confirmed quiescent galaxy at the center.

To evaluate the significance of this over-density of massive galaxies, we make a density plot in Figure \ref{fig:2}. This plot is obtained by the Gaussian kernel density estimation method. The galaxy distribution in the entire COSMOS field is smoothed over $30''$ (200 physical kpc at $z=4.5$, which roughly corresponds to the scale of this over-density region). Our sample here is complete for SFGs but not for QGs \citep{Weaver_2022complete}, and we may be missing low-mass QGs if any. SFGs significantly outnumber QGs at these redshifts and missing low-mass QGs will not strongly affect the density field. We find that our galaxy is located in a highly significant ($12\sigma$) over-density region, and this structure turns out to be the densest region of massive galaxies at $4.4<z_\mathrm{phot}<4.7$ in the COSMOS field. On the other hand, if the galaxy distribution is smoothed over $2.5'$ (1 physical Mpc at $z=4.5$), the region is a less significant ($2.2\sigma$) over-density. These kernel density does not follow the Gaussian distribution, and the absolute significance is not very important. We also estimate the excess of galaxy density using the 5th Nearest Neighbor method, which yields a density distribution more similar to Gaussian and obtains $180\sigma$. The group still remains the densest region in the entire COSMOS field. Thus, the group is very compact. For comparison, we indicate in Figure \ref{fig:2} a massive proto-cluster at $z\sim4.57$ with the blue circle \citep{Lemaux_2018}. As can be seen, the proto-cluster is much more extended than our group and on the scale of $\sim \mathrm{Mpc}$. The structure and size of our group is really different from typical proto-clusters at this high redshift (\citealp{Overzier_2016} for a review). 

\begin{deluxetable}{lccc}
\tablecaption{Coordinates (J2000) and VISTA/$K_s$-band magnitudes of massive galaxy group members  \label{tab:RADEC}}
\tablehead{
\colhead{ID} & \colhead{R.A.} & \colhead{Dec.} & \colhead{$m_{K_s}\,\mathrm{[mag]}$}
}
\colnumbers
\startdata 
{1048346} & $\mathrm{10^h02^m27^s.099}$ & $\mathrm{+02^\circ 24' 41''.607}$ & 23.45\\
{1046859} & $\mathrm{10^h02^m26^s.337}$ & $\mathrm{+02^\circ 24' 38''.491}$ & 24.85\\
{1048201} & $\mathrm{10^h02^m27^s.838}$ & $\mathrm{+02^\circ 24' 42''.755}$ & 25.13\\
{1041229} & $\mathrm{10^h02^m26^s.318}$ & $\mathrm{+02^\circ 24' 19''.907}$ & 24.72\\
\hline
{1047519} & $\mathrm{10^h02^m27^s.091}$ & $\mathrm{+02^\circ 24' 39''.672}$ & 23.16\\
\enddata

\end{deluxetable}

\begin{deluxetable*}{lcccccccc}
\tablecaption{Inferred physical properties of massive galaxy group members \label{tab:group}}
\tablehead{
\colhead{ID} & \colhead{$z_\mathrm{phot}$} & \colhead{$z_\mathrm{spec}$} & \colhead{$\theta_{\mathrm{sep}}$}&  \colhead{$\theta_{\mathrm{sep}}$} & \colhead{$\log{M_*/M_\odot}$} & \colhead{$SFR_\mathrm{SED}$} & \colhead{$\log{M_h/M_\odot}$} & \colhead{$r_{200}$} \\
\colhead{} & \colhead{} & \colhead{} & \colhead{[$''$]} & \colhead{[$\mathrm{kpc}$]} & \colhead{}& \colhead{[$M_\odot\,\mathrm{yr^{-1}}$]} & \colhead{} & \colhead{[kpc]} 
}
\colnumbers
\startdata 
{1048346} & $4.535^{+0.168}_{-0.105}$ & -- & 1.94 & 12.8 & $10.10\pm 0.13$ & $87.1^{+19.0}_{-28.2}$ & $11.91^{+0.76}_{-0.63}$ & $51.8^{+40.5}_{-19.9}$\\
{1046859} & $4.630^{+0.118}_{-0.114}$  & -- & 11.36 & 75.0 & $10.04\pm 0.14$ & $32.7^{+21.7}_{-17.8}$ & $11.88^{+0.76}_{-0.64}$ & $50.5^{+39.5}_{-19.6}$\\
{1048201} & $4.570^{+0.077}_{-0.075}$  & -- & 11.61 & 76.4 & $9.66\pm 0.09$ & $5.48^{+2.18}_{-1.14}$ & $11.66^{+0.70}_{-0.59}$ & $42.7^{+30.1}_{-15.4}$\\
{1041229} & $4.693^{+0.103}_{-0.104}$  & -- & 22.91 & 150.8 & $9.77\pm 0.18$ & $56.0^{+14.8}_{-16.1}$ & $11.72^{+0.77}_{-0.62}$ & $44.8^{+35.6}_{-16.8}$\\
\hline
{1047519\tablenotemark{a}} & $4.654^{+0.134}_{-0.104}$ & $4.5313\pm 0.0005$  & -- & -- &  $10.71\pm 0.04$ & $15.1^{+6.5}_{-5.2}$ & $12.26^{+0.74}_{-0.63}$ & $67.5^{+51.3}_{-25.8}$\\
\enddata

\tablecomments{
\begin{enumerate}
    \setlength{\itemsep}{-2pt}
    \item Object ID of member galaxies from the COSMOS2020 \textsc{Classic} catalog \citep{Weaver_2022}.
    \item Photometric redshift estimated from \texttt{MIZUKI} \citep{Tanaka_2015}.
    \item Spectroscopic redshift of COSMOS-1047519.
    \item Angular separation between COSMOS-1047519 and SFGs.
    \item Projected physical distance between COSMOS-1047519 and SFGs.
    \item Inferred stellar mass from \texttt{prospector}.
    \item Inferred star formation rate from \texttt{prospector}.
    \item The halo mass estimated from the stellar-to-halo mass relation (SHMR: \citealp{Shuntov_2022}).
    \item Virial radius estimated from the halo mass of each galaxy.
\end{enumerate}
\tablenotetext{a}{The physical properties of the QG are obtained from \texttt{prospector} using delayed tau-model.}
}
\end{deluxetable*}

\subsection{Halo mass and virial radius}
While we have not spectroscopically confirmed the redshifts of the SFGs yet, we make an attempt to infer the halo mass of each member and see if they can be physically associated. We use the stellar-to-halo mass relation (SHMR) observed in the COSMOS field at $4.5<z<5.5$ \citep{Shuntov_2022} to estimate the halo mass from the stellar mass based on \texttt{prospector} assuming that each member is a central galaxy. Estimated values are summarized in Table \ref{tab:group} (8). The most massive member (quiescent galaxy) has $M_h\sim 10^{12.3}\,M_\odot$, which is slightly smaller than the galaxy group at $z\gtrsim3$ from previous research ($M_h\gtrsim 10^{13}\,M_\odot$: e.g., \citealp{Miller_2018,Hill_2020,Daddi_2021}). Next, we translate the halo mass into the virial radius. Here, we use the radius $r_{200}$ \citep{Carlberg_1997} as the virial radius. Table \ref{tab:group} (9) summarizes the virial radii of the member candidates. The dotted circles in the inset of Figure \ref{fig:2} show the virial radii of the member candidates significantly overlap. Two SFGs are just outside of the virial radius of the quiescent galaxy in Figure \ref{fig:2}, but they are actually consistent with being inside the virial radius within the uncertainties (see Table \ref{tab:group}). The right bottom galaxy is outside of it, but there is intrinsic scatter in the SHMR. These 3\,-\,4 galaxies likely form a group with the quiescent galaxy at the center. Spectroscopic confirmations of these member candidates are required for further characterization of the group environment.

\subsection{Possible formation scenarios of the QG}
As discussed earlier, COSMOS-1047519 is quenched very recently (or is being quenched currently). The group is thus an ideal object to examine the role of environments, especially for quenching. From the discussion in Sections \ref{subsec:previous} and \ref{subsec:AGN}, the quiescent galaxy experienced intense star formation ($\sim 200\mathrm{-}400\,M_\odot\,\mathrm{yr^{-1}}$) and then rapid quenching started at $z\sim 5$. The peak SFR and quenching timescale from the SFH are similar to the SMG’s SFR and gas depletion timescale at $z\sim5$, respectively. Also, the $K$-band spectrum of the quiescent galaxy shows the excess of [O\,{\footnotesize II}] emission compared to the model spectrum from \texttt{prospector}, and this galaxy may harbor a low-luminosity AGN. Combining these points, the star formation and AGN activities induced by the galaxy-galaxy interactions and mergers drive the quenching of this quiescent galaxy. For example, observational and theoretical studies argue that minor mergers and galaxy interactions increase star formation efficiency and AGN activity (e.g., \citealp{Hopkins2008,Brodwin2013,Gomez-Guijarro2018,Coogan2018}). Also, previous studies confirm that gas mass fractions of QGs are significantly lower than those of SFGs (e.g., \citealp{Sargent2015,Bezanson2019,Williams2021,2021Natur.597..485W, Suzuki_2022}). Thus, it is possible that the quick gas depletion due to starburst is responsible for quenching. 

\cite{Shimakawa_2018} suggest that group/cluster environments have evolved stellar populations, and AGN feedback suppresses the star formation of member galaxies at high redshift ($2<z<3$). \cite{Chiang_2017} argue that significant group- or cluster-sized cores (with the scale of $r_{200}$) would be the first region to show the galaxy quenching. Our group candidate has a comparable scale of these cores, and the quenching observed in this group environment at $z>4$ is probably consistent with the core-scale quenching. Our work suggests this core-scale quenching has already occurred at $z>4$ by gas consumption due to the starburst and/or AGN feedback. Furthermore, recent James Webb Space Telescope (JWST) observations confirm compact galaxy groups at $5<z<8$ (e.g., \citealp{Sun2023,Jin2023,Helton2023,Morishita2023,Hashimoto2023}). Member galaxies are mostly low-mass SFGs, but \cite{Helton2023} show that more massive galaxies with larger SFR populate in the over-density region compared to the field. 
Therefore, the formation of galaxy groups with a QG at $z=4.5$ could result from the evolution of such very high-redshift compact galaxy groups.

\section{Conclusion} \label{sec:Concl}
We present the spectroscopic confirmation of a massive quiescent galaxy at $z=4.53$ based on the Keck/MOSFIRE observation. We confirm a very large stellar mass and a very low star formation rate from the extensive SED fitting using both the photometry and spectrum. The estimated SFR is more than 1 dex below the star-forming main sequence at $z=4.5$. The star formation history of the galaxy inferred from the SED fitting indicates rapid quenching with a timescale of $\sim 100\,\mathrm{Myr}$ from $z\sim 5$.

To constrain physical drivers of the starburst and subsequent quenching, we compare the properties of this object with those of quiescent galaxies from the literature. The galaxy is one of the youngest quiescent galaxies from the formation redshift; this is an interesting object to understand the physical processes responsible for the suppression of star formation. The fact that the quenching timescale of the galaxy is comparable to the gas depletion timescale of SMGs at similar redshift indicates that the quenching is simply due to gas consumption. Also, the observational excess of [O\,{\footnotesize II}] indicates the existence of low-luminosity AGN. In addition, the galaxy is found to be located in the group environment including a companion galaxy at $2''$ (13 physical kpc). These findings suggest that gas depletion due to the starburst and/or AGN feedback triggered by galaxy-galaxy interactions or mergers may be responsible for quenching. 

In Figure \ref{fig:formz}, we can clearly say that we are approaching the formation epoch of the massive quiescent galaxies. This is an exciting redshift regime to directly unveil the quenching physics. Recent JWST observations confirm quiescent galaxies at these redshifts \citep{carnall_2023} as well as a large number of the quiescent galaxy candidates \citep{Valentino_2023,Carnall_2023_photo,long_2023}. Upcoming imaging and spectroscopic observations will place a more stringent constraint on the physical mechanism of quenching and detailed star formation history in the near future.

\begin{acknowledgments}
  We thank the anonymous referee for constructive comments. This study was supported by JSPS KAKENHI Grant Number JP21K03622 (MO). PFW acknowledges funding through the National Science and Technology Council grant 111-2112-M-002-048-MY2 and National Taiwan University grant NTU 112L7318 and NTU 112L7439. The data presented herein were obtained at the W. M. Keck Observatory, which is operated as a scientific partnership among the California Institute of Technology, the University of California and the National Aeronautics and Space Administration. The Observatory was made possible by the generous financial support of the W. M. Keck Foundation.
  The authors wish to recognize and acknowledge the very significant cultural role and reverence that the summit of Maunakea has always had within the indigenous Hawaiian community.  We are most fortunate to have the opportunity to conduct observations from this mountain.
  Based on observations collected at the European Southern Observatory under ESO programme ID 179.A-2005 and on data products produced by CALET and the Cambridge Astronomy Survey Unit on behalf of the UltraVISTA consortium.
\end{acknowledgments}

%

\vspace{5mm}
\facilities{Keck:I (MOSFIRE)}


\software{Astropy \citep{2013A&A...558A..33A,2018AJ....156..123A},
          Matplotlib \citep{Hunter:2007},
          MOSFIRE DRP,
          \texttt{MIZUKI} \citep{Tanaka_2015},
          numpy \citep{harris2020array},
          \texttt{prospector} \citep{Johnson_2021},
          Python-fsps \citep{Conroy_2009,Conroy_2010},
          \texttt{emcee} \citep{2013PASP..125..306F}
          }



\bibliography{sample631}{}
\bibliographystyle{aasjournal}



\end{document}